\documentclass[a4paper]{article}
\usepackage{epsfig,latexsym,amsmath}

\newcommand{\bd}{\ensuremath{[B/D]}}

\newcommand{\var}[2][]{\mathrm{Var}_{#1}\left(#2\right)}
\newcommand{\cov}[3][]{\mathrm{Cov}_{#1}\left(#2,#3\right)}
\newcommand{\ex}[2][]{\mathrm{E}_{#1}\left(#2\right)}

\newcommand{\size}[2][d]{\mathrm{Size}_{#1}\left(#2\right)}
\newcommand{\sizeratio}[2][d]{\mathrm{Sr}_{#1}\left(#2\right)}
\newcommand{\proj}[2][D]{\mathrm{P}_{[#1]}^{\{#2\}}}
\newcommand{\transf}[2][D]{\mathrm{T}_{[#1]}^{\{#2\}}}
\newcommand{\ci}[3][]{#2\amalg#3/#1}
\newcommand{\trace}[1]{\mathrm{Tr}(#1)}
\newcommand{\trans}{\mbox{}^\mathrm{T}}
\newcommand{\inv}{\mbox{}^{\dagger}}

\newcommand{\id}{\mathrm{I}}

\newcommand{\mupad}{\textsf{\textsl{MuPAD}}}
\newcommand{\be}{\begin{eqnarray}}
\newcommand{\ee}{\end{eqnarray}}
\newcommand{\beq}{\begin{equation}}
\newcommand{\eeq}{\end{equation}}
\newcommand{\comment}[1]{}

\newtheorem{thm}{Theorem}
\newtheorem{lemma}{Lemma}

\newtheorem{defn}{Definition}

\newenvironment{proof}{\par\noindent\textsf{Proof}\par}{\hfill $\Box$\par\vspace{0.1in}\par}

\title{Local computation of influence propagation through Bayes linear
  belief networks}
\author{D.~J.~Wilkinson%
\thanks{E-mail: \texttt{d.j.wilkinson@newcastle.ac.uk}, 
WWW URL: \texttt{http://www.ncl.ac.uk/$\sim$ndjw1/}}}
\date{\today}

\begin{document}

\maketitle

\begin{abstract}
In recent years there has been interest in the theory of local
computation over probabilistic Bayesian graphical models. In this paper, local
computation over Bayes linear belief networks is shown to be amenable
to a similar approach. However, the linear structure offers many
simplifications and advantages relative to more complex models, and
these are examined with reference to some illustrative examples. 
\end{abstract}


\section{Introduction}
Conditional independence graphs are of vital importance in the
structuring, understanding and computing of high dimensional complex
statistical models. For a review of early work in this area, see
\cite{lslcprob}, the references and the discussion, and also
\cite{apdprop}.  The above mentioned work is concerned
with updating in discrete probability networks. For a discussion of
updating in networks with continuous random variables, see
\cite{sllprop}, for example. For a general overview
of the theory of graphical models, see \cite{sllbook}.

Also relevant to this paper is the work on graphical Gaussian
models. \cite{sllbook}, \cite{whit} and \cite{tpshk}
discuss the properties of  such models.
\cite{ntbayesnet} examine
data propagation through a graphical Gaussian network, and apply their
results to a dynamic linear model (DLM).
 Here, the aim is to link the
theory of local computation over graphical Gaussian networks to the
Bayes linear framework for 
subjective statistical inference, and the many interpretive and
diagnostic features associated with that methodology, in particular.

\section{Bayes linear methods}
\subsection{Overview}
In this paper, a Bayes linear approach is taken to subjective statistical inference,
making expectation (rather than probability) primitive. An overview of
the methodology is given in \cite{fgcross}.
The foundations of the theory are quite general, and are outlined in
the context of second-order exchangeability in \cite{mgrevexch},
and discussed for more general situations in \cite{mgpriorinf}.
Bayes
linear methods may be used in order to learn about any quantities of
interest, provided only that a mean and variance specification is made
for all relevant quantities, and a specification for the covariance
between all pairs of quantities is made. No distributional assumptions
are necessary.
There are many interpretive and
diagnostic features of the Bayes linear methodology. These are
discussed with reference to \bd\ (the Bayes linear computer
programming language) in \cite{gwblincomp}. 

\subsection{Bayes linear conditional independence}
Conventional graphical models are defined via strict probabilistic
conditional independence \cite{apdci}. However, as \cite{jqsid}
demonstrates,  all that is actually required is a tertiary operator
$\ci[\cdot]{\cdot}{\cdot}$ satisfying some simple properties. Any
relation satisfying these properties is known as a \emph{generalised
  conditional independence} relation. Bayes
linear graphical models are based on what \cite{jsstatgraph}
refers to as \emph{weak conditional independence}. In this paper, the
relation will be referred to as \emph{adjusted orthogonality}, in
order to emphasise the linear structure underlying the relation.

 Bayes
linear graphical models based upon the concept of adjusted
orthogonality are described in \cite{mginfl}. For completeness,
and to introduce some notation useful in the context of local
computation, the most important elements of the methodology are
summarised here, and the precise form of the adjusted orthogonality
relation is defined. 

For vectors of random quantities, $X$ and $Y$, define
$\cov{X}{Y}=\ex{XY\trans}-\ex{X}\ex{Y}\trans$ and
$\var{X}=\cov{X}{X}$. Also, for any matrix, $A$, $A\inv$ represents
the Moore-Penrose generalised inverse of $A$.
\begin{defn}
For all vectors of random quantities $B$ and $D$, define
\begin{align}
\proj[D]{B} =& \cov{B}{D}\var{D}\inv \\
\transf[D]{B} =& \proj{B}\proj[B]{D}
\end{align}
\end{defn}
These represent the fundamental operators of the
Bayes linear methodology. $\proj[D]{B}$ is the operator which updates
the expectation vector for $B$ based on the observation of $D$, and
$\transf[D]{B}$ updates the variance matrix for $B$ based on
observation of $D$. Local computation over Bayes linear graphical
models is made possible by local computation of these operators. 
\begin{defn}
For all vectors of random quantities $B$, $C$ and $D$, define
\begin{align}
\ex[D]{B} =& \ex{B}+\proj[D]{B}[D-\ex{D}] \label{eq:exbd}\\ 
\cov[D]{B}{C} =& \cov{B-\ex[D]{B}}{C-\ex[D]{C}} \label{eq:covdbc}
\end{align}
\end{defn}
$\ex[D]{B}$ is the \emph{expectation for $B$ adjusted by
  $D$}. It represents the linear combination of a constant and the
components of $D$ \emph{closest to} $B$ in the sense of expected
squared loss. It corresponds to $\ex{B|D}$ when $B$ and $D$ are
jointly multivariate normal. $\cov[D]{B}{C}$ is the 
\emph{covariance between $B$ and $C$ adjusted by $D$}, and represents
the covariance between $B$ and $C$ given
observation of $D$. It corresponds to $\cov{B}{C|D}$ when $B$, $C$ and
$D$ are jointly multivariate normal.
\begin{lemma}
For all vectors of random quantities $B$, $C$ and $D$
\begin{align}
\cov[D]{B}{C} =& \cov{B}{C}-\cov{B}{D}\proj[D]{C}\trans \label{eq:covdbc2}\\
\var[D]{B} =& (\id-\transf[D]{B})\var{B}\label{eq:vardb2}
\end{align}
\end{lemma}
\begin{proof}
Substituting \eqref{eq:exbd} into \eqref{eq:covdbc} we get
\begin{align}
\cov[D]{B}{C} =& \cov{B}{C-\ex[D]{C}} \\
=& \cov{B}{C-\proj[D]{C}D}
\end{align}
which gives \eqref{eq:covdbc2}, and replacing $C$ by $B$ gives
\eqref{eq:vardb2}. 
\end{proof}
Note that \eqref{eq:vardb2} shows that $\transf[D]{B}$ is responsible
for the updating of variance matrices. Adjusted orthogonality is now defined.
\begin{defn}
\label{def:ci}
For random vectors $B$, $C$ and $D$
\begin{align}
\ci[D]{B}{C} \iff& \cov[D]{B}{C}=0
\end{align}
\end{defn}
\cite{mginfl} shows that this relation does indeed define a
generalised conditional independence property, and hence that all the
usual properties of graphical models based upon such a relation hold. 

\subsection{Bayes linear graphical models}
\cite{mginfl} defines a Bayes linear influence
diagram based upon the adjusted orthogonality relation. 
\cite{gfsbeer} illustrate the use of Bayes linear influence diagrams in
a multivariate forecasting problem.
 Relevant graph
theoretic concepts can be found in the appendix of
\cite{apdsllhyper}. The terms \emph{moral graph} and \emph{junction
  tree} are explained in \cite{sdlc}.
Briefly, an undirected moral graph is formed from a directed acyclic
graph by \emph{marrying} all pairs of parents of each node, by adding an arc
between them, and
then dropping arrows from all arcs. A junction tree is the \emph{tree} of
\emph{cliques} of a \emph{triangulated} moral graph. A tree is a graph
without any cycles. A graph is
triangulated if no cycle of length at least four is without a chord. A
clique is a maximally connected subset of a triangulated graph. 

 In this paper,
attention will focus on undirected graphs. An undirected graph
consists of a collection of nodes $B=\{B_i|1\leq i\leq n\}$ for some
$n$, together with a collection of undirected arcs. Every pair
of nodes, $\{B_i,B_j\}$ is joined by an undirected arc unless
$\ci[B\backslash\{B_i,B_j\}]{B_i}{B_j}$. Here, the standard
set theory notation, $B\backslash A$ is used to mean the set of elements of $B$
which are not in $A$. An
undirected graph may be obtained from a Bayes linear influence diagram
by forming  the
moral graph of the influence diagram in the usual way. 

In fact, \emph{local computation} (the computation of global
influences of particular nodes of the graph, using only information
local to adjacent nodes) requires that the undirected graph
representing the conditional independence structure is a
tree.  This
tree may be formed  as the junction
  tree  of a triangulated moral graph, or better, by grouping together
related variables ``by hand'' in order to get a tree structure for the
graph. For the rest of this paper, it will be assumed that the model
of interest is represented by an undirected tree defined via adjusted
orthogonality.  

\section{Local computation on Bayes linear graphical models}
\subsection{Transforms for adjusted orthogonal belief structures}
\begin{lemma}
\label{lem:covbc}
If $B$, $C$ and $D$ are random vectors such that $\ci[D]{B}{C}$, then
\begin{equation}
\cov{B}{C}=\cov{B}{D}\proj[D]{C}\trans
\end{equation}
\end{lemma}
This follows immediately from Definition \ref{def:ci} and \eqref{eq:covdbc2}.
\begin{lemma}
\label{lem:covxyz}
If $X$, $Y$ and $Z$ are random vectors such that $\ci[Y]{X}{Z}$, then
\begin{align}
\cov[X]{Y}{Z} =& (\id-\transf[X]{Y})\cov{Y}{Z}
\end{align}
\end{lemma}
\begin{proof}
From \eqref{eq:covdbc2}
\begin{align}
\cov[X]{Y}{Z} =& \cov{Y}{Z}-\cov{Y}{X}\var{X}\inv\cov{X}{Z} \\
=& \cov{Y}{Z} - \proj[X]{Y}\cov{X}{Y}\proj[Y]{Z}\trans \quad
\mathrm{by\ Lemma\ \ref{lem:covbc}}
\end{align}
and the result follows.
\end{proof}
\begin{thm}
\label{thm:lc}
If $X$, $Y$ and $Z$ are random vectors such that $\ci[Y]{X}{Z}$, then
\begin{align}
\proj[X]{Z} =& \proj[Y]{Z}\proj[X]{Y} \label{eq:projxz} \\
\transf[X]{Z} =& \proj[Y]{Z}\transf[X]{Y}\proj[Z]{Y} \label{eq:transfxz}
\end{align}
\end{thm}
\begin{proof}
\begin{align}
\proj[X]{Z} =& \cov{Z}{X}\var{X}\inv \\
=& \cov{Z}{Y}\proj[Y]{X}\trans\var{X}\inv \quad \mathrm{by\ Lemma\
  \ref{lem:covbc}}
\end{align}
which gives \eqref{eq:projxz}. Also
\begin{align}
\transf[X]{Z} =& \proj[X]{Z}\proj[Z]{X} \\
=& \cov{Z}{X}\var{X}\inv\cov{X}{Z}\var{Z}\inv \\
=&
\cov{Z}{Y}\proj[Y]{X}\trans\var{X}\inv\cov{X}{Y}\proj[Y]{Z}\trans\var{Z}\inv
\end{align}
which gives \eqref{eq:transfxz}.
\end{proof}
Theorem \ref{thm:lc} contains the two key results which allow local
computation over Bayes linear belief networks. 

\subsection{Local computation on trees}
\begin{figure}
\epsfig{file=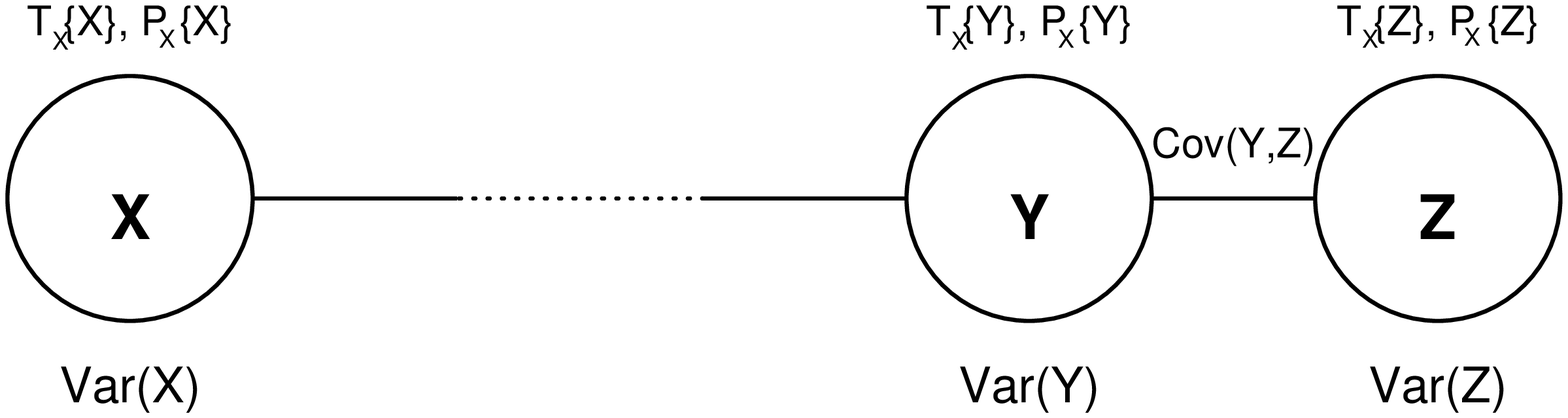,width=6in}
\caption{Local computation along a path}
\label{fig:path}
\end{figure}
The
implications of Theorem \ref{thm:lc} to Bayes linear trees should be
clear from examination of Figure \ref{fig:path}.
To examine the effect of observing node $X$, it is sufficient to
 compute the operators $\proj[X]{Z}$ and
$\transf[X]{Z}$ for every node, $Z$ on the
graph, since these operators contain all necessary
information about the adjustment of $Z$ by $X$.
There is a unique path
from $X$ to $Z$ which is shown in
Figure \ref{fig:path}. The direct predecessor of $Z$ is denoted by
$Y$. Note further that it is a property of the graph that
$\ci[Y]{X}{Z}$. Further,  by Theorem \ref{thm:lc}, the
transforms $\proj[X]{Z}$ and $\transf[X]{Z}$ can be computed using
$\proj[X]{Y}$ and $\transf[X]{Y}$ together with information local to
nodes $Y$ and $Z$. This provides a recursive method for the
calculation of the transforms, which leads to the algorithm for the
propagation of transforms throughout the tree, which is described in
the next section.

\subsection{Algorithm for transform propagation}
Consider a tree with nodes $B=\{B_1,\ldots,B_n\}$ for some
$n$. Each node, $B_i$, represents a vector of random quantities. It
also has an edge set $G$,
where each $g\in G$ is of the form $g=\{B_k,B_l\}$ for some $k,l$. The
resulting tree should represent a conditional independence graph over
the random variables in question. It
is assumed that each node, $B_i$ has an expectation vector $E_{B(i)}=\ex{B_i}$
and variance matrix $V_{B(i)}=\var{B_i}$ associated with it. It is further
assumed that each edge, $\{B_k,B_l\}$ has the covariance matrix,
$C_{B(k),B(l)}=\cov{B_k}{B_l}$ associated with it. This is the only
information required in order to carry out Bayes linear local
computation over such structures.

Now consider the effect of
adjustment by the vector $X$, which consists of some or all of the
components of node $B_j$ for some $j$. Then, starting with node $B_j$, 
calculate and store $T_{B(j)}=\transf[X]{B_j}$ and
$P_{B(j)}=\proj[X]{B_j}$. Then, for each node
$B_k\in b(B_j)\equiv\left\{B_i|\{B_i,B_j\}\in G\right\}$ calculate and
store $T_{B(k)}$ and $P_{B(k)}$, then for each node
$B_l\in b(B_k)\backslash B_j$, do the same, using Theorem \ref{thm:lc} to
calculate
\begin{align}
P_{B(l)} =& \proj[B_k]{B_l}P_{B(k)}  \\
T_{B(l)} =& \proj[B_k]{B_l}T_{B(k)}\proj[B_l]{B_k}
\end{align}
In this way, recursively step outward through the tree, at
each stage computing and storing the transforms using the transforms
from the predecessor and the variance and covariance information over and
between the current node and its predecessor. 

Once this process is completed, associated with every node, $B_i\in
B$, there are matrices 
$T_{B(i)}=\transf[X]{B_i}$ and $P_{B(i)}=\proj[X]{B_i}$. These
operators represent all information about the adjustment of the
structure by $X$. Note however, that $X$ has not yet been observed, and
that expectations, variances and covariances associated
with nodes and edges have not been updated. 

It is a crucial part of the Bayes linear methodology that \emph{a
  priori} analysis of the model takes place, and that the expected
influence of potential observables is examined. Examination of the
eigen structure of the belief transforms associated with nodes of
particular interest is the key to understanding the structure of the
model, and the benefits of observing particular nodes. It is important
from a design perspective that such analyses can take place before any
observations are made. See \cite{mgcomp} for a more complete
discussion of such issues, and \cite{gwblincomp} for a
discussion of the technical issues it raises.

\subsection{Updating of expectation and covariance structures after
  observation}
After observation of $X=x$, updating of the expectation,
variance and covariance structure over the tree is required. Start at
node $B_j$ and  calculate
\begin{align}
E_{B(j)}^\prime =& E_{B(j)} + P_{B(j)}\left(x-\ex{X}\right) \\
V_{B(j)}^\prime =& (\id-T_{B(j)})V_{B(j)}
\end{align}
(using \ref{eq:vardb2}).
Replace $E_{B(j)}$ by $E_{B(j)}^\prime$ and $V_{B(j)}$ by
$V_{B(j)}^\prime$. Then for each $B_k\in b(B_j)$ do the same, and also
update the arc between $B_j$ and $B_k$ by calculating
\begin{align}
C_{B(j),B(k)}^\prime =& (\id-T_{B(j)})C_{B(j),B(k)}
\end{align}
(using Lemma \ref{lem:covxyz}), and replacing $C_{B(j),B(k)}$ by
$C_{B(j),B(k)}^\prime$. Again, step outwards through the tree, updating
nodes and edges using the transforms previously calculated.

\subsection{Pruning the tree}
Once the expectations, variances and covariances over the structure
have been updated, the tree should be pruned. If the adjusting node
was completely observed (\emph{i.e.} $X=B_j$), then $B_j$ should be
removed from $B$, and $G$ should have any arcs involving $B_j$
removed. Further,  leaf nodes and their edges may always
be dropped without affecting the conditional independence structure of
the graph. This is important if a leaf node is partially observed and
the remaining variables are unobservable and of little diagnostic
interest, since it means that the whole node may be dropped after
observation of its observable components.

If a non-leaf node is
partially observed, or a leaf node is observed, but its remaining
components are observable or of interest, then the graph itself should
remain unaffected, but the expectation, variance and covariance
matrices associated with the node and its arcs should have the
observed (and hence redundant) rows and columns removed (for reasons
of efficiency --- use of the Moore-Penrose generalised inverse ensures
that no problems will arise if observed variables are left in the system).

\subsection{Sequential adjustment}
As data becomes available on various nodes, it should be incorporated
into the tree one node at a time. For each node with observations, the
transforms should be computed, and then the beliefs
updated in a sequential fashion. The fact that such sequential
updating provides a coherent method of adjustment is demonstrated in
\cite{mgadjbel}. 

\subsection{Local computation of diagnostics}
Diagnostics for Bayes linear adjustments are a crucial part of the
methodology, and are discussed in \cite{mgtraj}. It follows that
for local computation over Bayes linear networks to be of practical
value, methodology must be developed for the local computation of
Bayes linear diagnostics such as the \emph{size}, \emph{expected size}
and \emph{bearing}  of
an adjustment.
The bearing represents the magnitude and direction of changes
in belief. The magnitude of the bearing, which indicates the magnitude
of changes in belief, is known as the \emph{size} of the
adjustment.

Consider the observation of data, $X=x$, and 
the partial bearing of the adjustment it
induces on some node, $B_p$. Before observation of $X$, record $E=E_{B(p)}$
and $V=V_{B(p)}$. Also calculate the Cholesky factor, $A$ of $V$, so
that $A$ is lower triangular, and $V=AA\trans$. Once the observed
value $X=x$ is known, propagate the revised expectations, variances
and covariances through the Bayes linear tree. The new value of
$E_{B(p)}$ will be denoted $E_{B(p)}^\prime$. Now the quantity
\begin{align}
E^\prime =& A\inv(E_{B(p)}^\prime-E)
\label{eq:bear}
\end{align}
represents the adjusted expectation for an orthonormal basis,
$F=A\inv(B_p-E)$ for $B_p$ 
with respect to the \emph{a priori} beliefs, $E$ and $V$.
Therefore, $E^\prime$ gives the coordinates of
 the \emph{bearing} of the adjustment with respect to that
basis.

The 
size of the partial adjustment is given by
\begin{align}
\size[x]{B_p} =& ||E^\prime||^2
\end{align}
where $||\cdot||$ represents the Euclidean norm. The expected size is
given by
\begin{align}
\ex{\size[X]{B_p}} =& \trace{T_{B(p)}}
\end{align}
and so the \emph{size ratio} for the adjustment (often of most
immediate interest) is given by
\begin{align}
\sizeratio[x]{B_p} =& \frac{||E^\prime||^2}{\trace{T_{B(p)}}}
\end{align}
A size ratio close to one indicates changes in belief close to what
would be expected. A size ratio smaller than one indicates changes in
belief of smaller magnitude than would have been anticipated \emph{a
priori}, and a size ratio bigger than one indicates changes in belief
of larger magnitude than would have been expected \emph{a
priori}. Informally, a size ratio bigger than 3 is often taken to
indicate a diagnostic warning of possible conflict between \emph{a
priori} belief specifications  and the observed data. 

Cumulative sizes and bearings may be calculated in exactly the same
way, simply by updating several times before computing
$E^\prime$. However, to calculate the expected size of the adjustment,
in order to compute the size ratio, the cumulative belief transform
must be recorded and updated at each stage, using the fact that
\begin{align}
\transf[X+Y]{B} =& \ \id - (\id - \transf[Y/]{B/})(\id - \transf[X]{B})
\end{align}
where $\transf[Y/]{B/}$ represents the partial transform for $B$ by
$Y$, with respect to the structure already adjusted by $X$. In other
words, $\id$ minus the transforms at each stage multiply together to
give $\id$ minus the cumulative transform. See \cite{mgcomp}
for a more complete discussion of the multiplicative properties of
belief transforms.

\subsection{Efficient computation for evidence from multiple nodes}
To adjust the tree given data at multiple nodes, it would be
inefficient to adjust the entire tree sequentially by each node in
turn, if the nodes in question are ``close together''. Here again,
ideas may be borrowed from theory for the updating of probabilistic expert
systems. It is possible to propagate transforms and projections from
each node for 
which adjustment is required, to a \emph{strong root}, and then
propagate transforms from the 
strong root out to the rest of the tree. A strong root is a node which
separates the nodes for which there is information, from as much as possible
of the rest of the tree. In practice, there are many ways in which one
can use the
strong root in order to control information flow through the tree. An
example of its use is given in Section \ref{sec:nse}.

\subsection{Geometric interpretation, and infinite collections}

In this paper, attention has focussed exclusively on finite
collections of quantities, and matrix representations for Bayes linear
operators. All of the theory has been developed from the perspective
of pushing matrix representations of linear operators around a
network. However, the Bayes linear methodology may be formulated and
developed from a purely geometric viewpoint, involving linear
operators on a (possibly infinite dimensional) Hilbert space. This is
not relevant to practical computer implementations of the theory and
algorithms -- hence the focus on matrix formulations in this
paper. However, from a conceptual viewpoint, it is very important,
since one sometimes has to deal, in principle, with infinite
collections of quantities, or probability measures over an infinite
partition. In fact, all of the theory for local computation over Bayes
linear belief networks developed in this paper is valid for the local
computation of Bayes linear operators on an arbitrary Hilbert
space. Consequently, the results may be interpreted geometrically, as
providing a method of pushing linear operators around a Bayes linear
Hilbert space network. A geometric form of Theorem \ref{thm:lc} is derived
and utilised in \cite{gwaeb}.

\section{Example: A dynamic linear model}
\begin{figure}
\epsfig{file=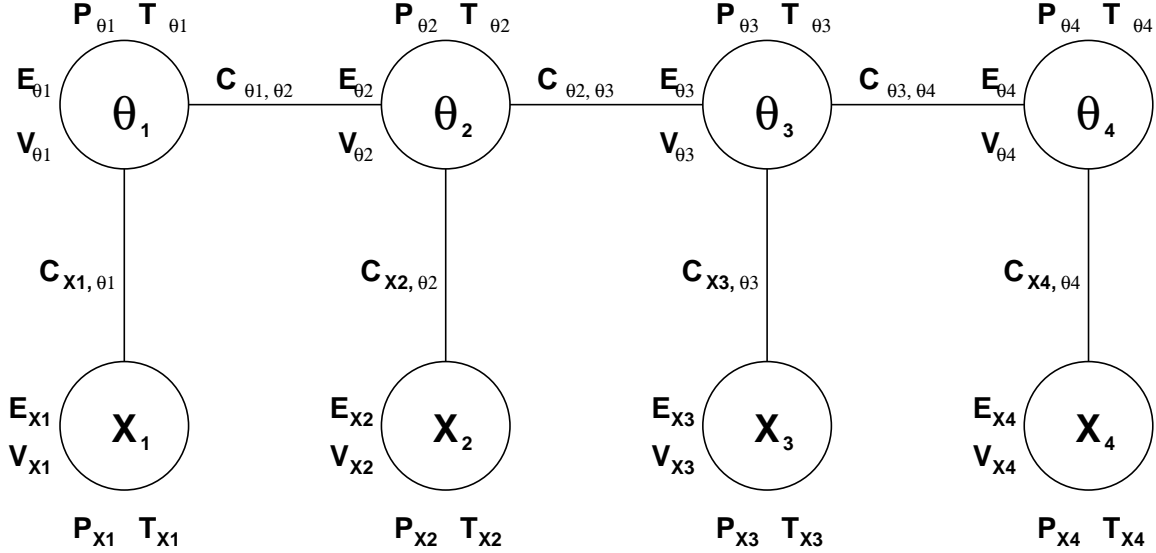,width=6in}
\caption{Tree for a dynamic linear model}
\label{fig:dlm}
\end{figure}
Figure \ref{fig:dlm} shows a Bayes linear graphical tree model for the
first four time points of a dynamic linear model. Local computation
will be illustrated using the
example model, beliefs and data from \cite{djwll}. Here,
$\forall t,\ \theta_t$ represents the vector $(M_t,N_t)\trans$ from
that paper.
The model takes the form
\begin{align}
X_t =& (1,0)\theta_t + \nu_t \\
\theta_t =& \left(\begin{array}{cc}
1 & 1\\
0 & 1
\end{array}\right)
\theta_{t-1} + \omega_t
\end{align}
where $\var{\theta_1}=diag(400,9)$, $\ex{\theta_1}=(20,0)\trans$,
$\ex{\nu_t}=0$, 
$\ex{\omega_t}=0,\ \var{\nu_t}=171,\ \var{\omega_t}=diag(4.75,0.36)\
\forall t$ and the $\nu_t$ and $\omega_t$ are uncorrelated. 

 First, the nodes and arcs shown in Figure \ref{fig:dlm}
are defined. Then the expectation and variance of each node is
calculated and associated with each node, and the covariances between
pairs of nodes joined by an arc is also computed, and associated with
the arc. All of the expectations variances and covariances are
determined by the model. For example, node $X_1$ has expectation
vector $(20)$ and variance matrix $(571)$ associated with it. Node
$\theta_1$ has expectation vector $(20,0)\trans$ and variance matrix
$diag(400,9)$ associated with it. The arc between $X_1$ and $\theta_1$
has associated with it the covariance matrix $(400,0)$. Note that
though the arc is undirected, the ``direction'' with respect to which
the covariance matrix is defined is important, and needs also to be
stored.

The effect of the observation of $X_1$ on
the tree structure will be examined, and the effect on the node
$\theta_4$ in particular, which 
has \emph{a priori} variance matrix $\left(\begin{array}{rr}
500.3 & 29.2 \\
29.2 & 10.1
\end{array}\right)$ associated with it. Before actual observation
of $X_1$, the belief transforms for the adjustment, may be computed
across the tree structure. The transforms are computed recursively,
in the following order.
\begin{xalignat}{3}
T_{X(1)}=&\left(\begin{array}{r}
1
\end{array}\right) &
T_{\theta(1)}=&\left(\begin{array}{rr}
0.7 & 0\\
0 & 0
\end{array}\right) &
T_{\theta(2)}=&\left(\begin{array}{rr}
0.692 & -0.692\\
0 & 0
\end{array}\right) \\
T_{\theta(3)}=&\left(\begin{array}{rr}
0.684 & -1.342\\
0 & 0
\end{array}\right) &
T_{\theta(4)}=&\left(\begin{array}{rr}
0.674 & -1.949\\
0 & 0
\end{array}\right) &
T_{X(4)}=&\left(\begin{array}{r}
0.417
\end{array}\right) \\
T_{X(3)}=&\left(\begin{array}{r}
0.453
\end{array}\right) &
T_{X(2)}=&\left(\begin{array}{r}
0.479
\end{array}\right) 
\end{xalignat}
The $P$ matrices are calculated similarly. In particular,
$P_{\theta(4)}=(0.7,0)\trans$. \emph{A priori} analysis of the belief
transforms is possible. For example, $\trace{T_{\theta(4)}}=0.674$,
indicating that observation of $X_1$ is expected to reduce overall
uncertainty about $\theta_4$ by a factor of $0.674$. This is also the
expected size of the bearing for the adjustment of $\theta_4$ by
$X_1$. 

Now, $X_1$ is observed to be $17$, and so the expectations, variances
and covariances may be updated across the structure. For example,
beliefs about node $\theta_4$ were updated so that
$E_{\theta(4)}=(17.9,0)\trans$ and $V_{\theta(4)}=\left(\begin{array}{rr}
220.0 & 29.2 \\
29.2 & 10.1
\end{array}\right)$ after propagation. Also, calculating the bearing
for the adjustment of $\theta_4$ by $X_1=17$, using \eqref{eq:bear}
gives $E^\prime = (-0.094,0.042)\trans$. Consequently, the size of the
adjustment is $0.011$ and the size ratio is $0.016$.

Once evidence from the observed value of $X_1$ has been taken into
account, the $X_1$ node, and the arc between $X_1$ and $\theta_1$ may
be dropped from the graph. Note also that $\theta_1$ then becomes an
unobservable leaf node, which may be of little interest, and so if
desired, the $\theta_1$ node, and the arc between $\theta_1$ and
$\theta_2$ may also be dropped. Observation of $X_2$ may now be
considered. Using the updated, pruned tree, projections and transforms
for the adjustment by $X_2$ may be calculated and propagated through
the tree. For example, the (partial) belief transform for the
adjustment of $\theta_4$ by $X_2$ is $\left(\begin{array}{rr}
0.46 & -0.87 \\
0.03 & -0.05
\end{array}\right)$. If cumulative diagnostics are of interest,
then it is necessary to calculate the cumulative belief transform,
$\left(\begin{array}{rr}
0.82 & -1.92 \\
0.01 & 0.00
\end{array}\right)$. This has trace $0.82$, and so the resolution for
the combined adjustment of $\theta_4$ by $X_1$ and $X_2$ is
0.82. Similarly, the expected size of the cumulative bearing is
$0.82$. $X_2$ is observed to be $22$. The new expectations, variances
and covariances may then be propagated through the tree. For example,
the new expectation vector for $\theta_4$ is $(19.95,0.13)\trans$. The
size of the cumulative bearing is $0.002$, giving a size ratio of
approximately $0.002$. Again, the tree may be pruned, and the whole
process may continue.

\section{Example: $n$-step exchangeable adjustments}
\label{sec:nse}
\begin{figure}[t]
\epsfig{file=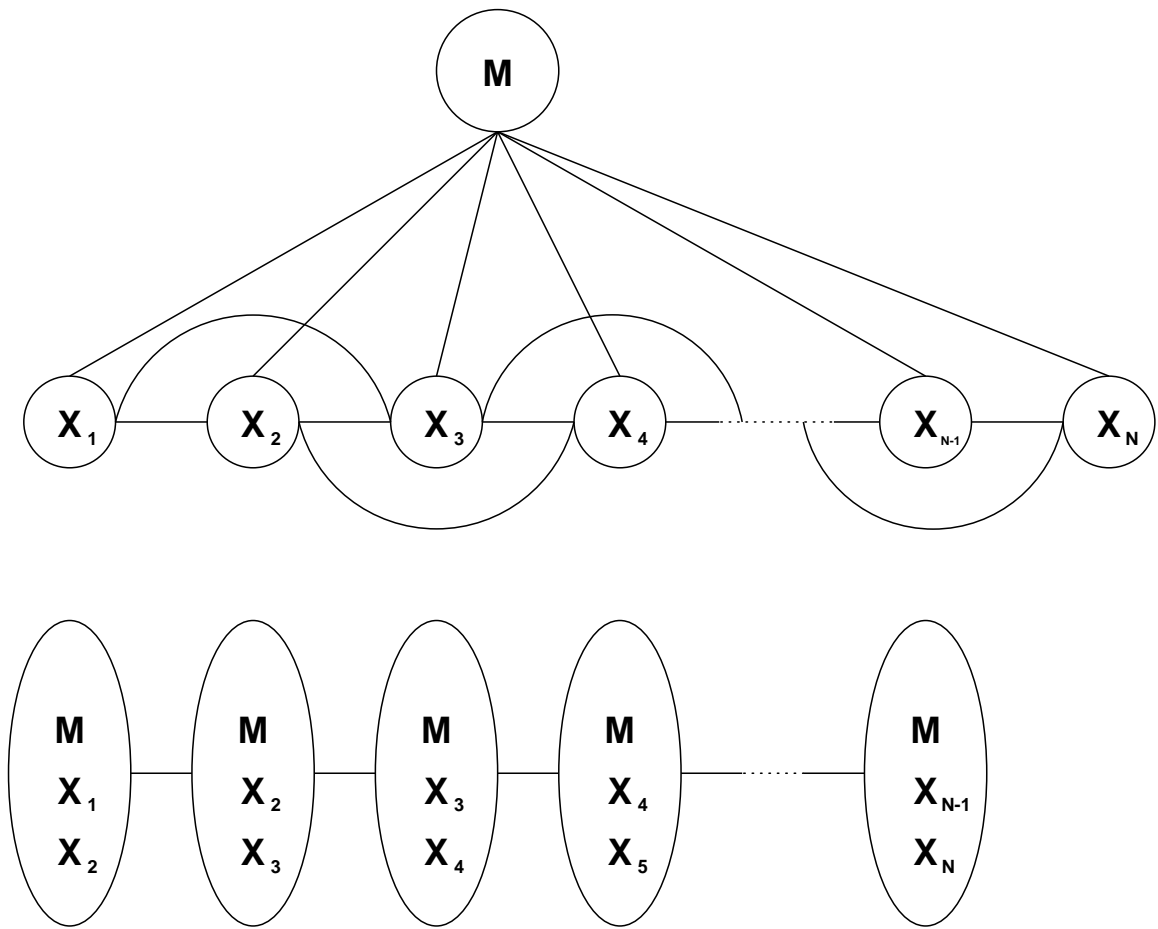,width=6in}
\caption{Graphical models for 3-step exchangeable quantities}
\label{fig:3se}
\end{figure}
An ordered collection of random quantities, $\{X_1,X_2,\ldots\}$ is
said to be (second-order) 
$n$-step exchangeable if (second-order) beliefs about the collection
remain invariant under an arbitrary translation or reflection of the
collection, and if the covariance between any two members of the
collection is fixed, provided only that they are a distance of at
least $n$ apart. Such quantities arise naturally in the context of
differenced time series \cite{djwll}. $n$-step exchangeable quantities may be
written in the form
\begin{align}
X_i =& M + R_i,\ \forall i
\end{align}
where the $R_i$ are a mean zero $n$-step exchangeable collection such
that the covariance between them is zero provided they are a distance
of at least $n$ apart. $M$ represents the underlying mean for the
collection, and $R_i$ represents the residual uncertainty which would
be left if the underlying mean became known.
Introduction of a mean
quantity helps to simplify a graphical model for an $n$-step
exchangeable collection. For example, Figure \ref{fig:3se} (top) shows an
undirected graphical model for a $3$-step exchangeable
collection. Note that without the introduction of the mean quantity,
$M$, all nodes on the graph would be joined, not just those a distance
of one and two apart. Figure \ref{fig:3se} (bottom) shows a
conditional independence graph for the same collection of quantities,
duplicated and grouped together so as to make the resulting graph
a tree. Note that each node contains $3$ quantities, and that there is
one less node than observables.

\begin{figure}[t]
\epsfig{file=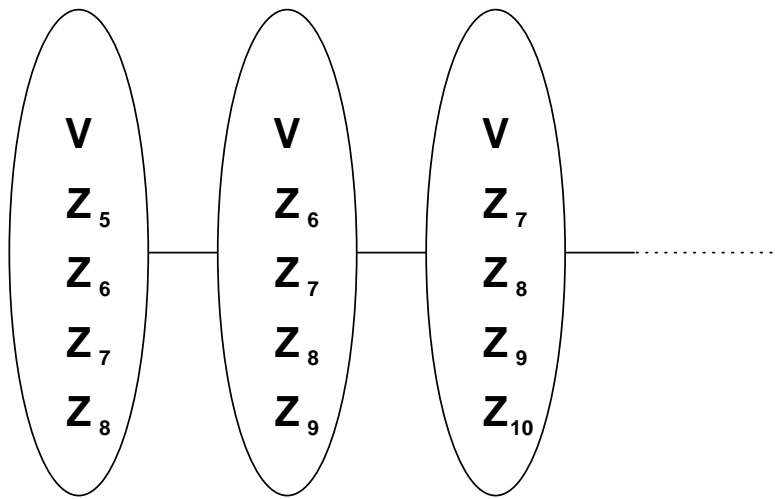,width=4in}
\caption{A graphical model for 5-step exchangeable vectors}
\label{fig:5se}
\end{figure}

In general, for a collection of $N$,
$n$-step exchangeable quantities, the variables can be grouped
together to obtain a simple chain graph, in the obvious way, so that
there are $N-n+2$ nodes, each containing $n$ quantities. The resulting
graph for 
$5$-step exchangeable quantities is shown in Figure \ref{fig:5se}
(with the first four nodes missing).

In \cite{djwll}, $3$-, $4$-, and $5$-step exchangeable
collections, $\{{X^{(1)}_3}^2,{X^{(1)}_4}^2,\ldots,\}$,
$\{{X^{(2)}_4}^2,{X^{(2)}_5}^2,\ldots,\}$ and
$\{{X^{(3)}_5}^2,\linebreak {X^{(3)}_6}^2,\ldots,\}$ are used in order to learn
about the quantities, $V_1$, $V_2$ and $V_3$, representing  the
variances underlying the DLM discussed in the previous section. Since
the observables sequences are $3$-, $4$-, and $5$-step exchangeable,
they may all be regarded as $5$-step exchangeable, and so Figure
\ref{fig:5se} represents a graphical model for the variables, where
$V=(V_1,V_2,V_3)\trans$, and $\forall i,\
Z_i=({X^{(1)}_i}^2,{X^{(2)}_i}^2,{X^{(3)}_i}^2)\trans$. Note that $V$
represents (a known linear function of) the mean of the $5$-step
exchangeable vectors, $Z_t$. Each node of the graph actually contains
$15$ quantities. For example, the first node shown contains $\{ 
V_1,V_2,V_3,{X^{(1)}_5}^2,{X^{(2)}_5}^2,{X^{(3)}_5}^2,
{X^{(1)}_6}^2,{X^{(2)}_6}^2,{X^{(3)}_6}^2,\linebreak
{X^{(1)}_7}^2,{X^{(2)}_7}^2,{X^{(3)}_7}^2
,{X^{(1)}_8}^2,{X^{(2)}_8}^2,{X^{(3)}_8}^2
\}$. Note that the fact that quantities are duplicated in other nodes
does not affect the analysis in any way. Observation of a particular
quantity in one node will reduce to zero the variance of that quantity
in any other node, as they will have a correlation of unity. Here,
the fact that  the Moore-Penrose generalised inverse is used in the
definition of the projections and transforms becomes
important. Updating for this  model may be locally computed over this
tree structure in the usual way. 

Suppose now that information for quantities $Z_5$ to $Z_9$ is to
become available simultaneously. This corresponds to information on
the first two nodes (and others, but this will be conveyed
automatically). The second node acts as a strong root for information
from the first two nodes. The transform for the first node may be
calculated using information on the first node, thus allowing
computation of the transform for the
second node given information on the first. Once the information from
the first node has been incorporated into the first two nodes, the
transform for the 
 second node  given information
from the first two nodes may be calculated, and the resulting
transform for the second node given information on the first two may
be used in order to propagate information to the rest of the tree. 

\section{Implementation considerations}
A test system for model building and computation over Bayes linear
belief networks has been developed by the author using the \mupad\
computer algebra system, described in
\cite{mupadp} and \cite{mupadb}. \mupad\ is a very high level
object-oriented mathematical programming language, with symbolic
computing capabilities, ideal for the rapid prototyping of
mathematical software and algorithms. The test system allows definition of
nodes and arcs of a tree, and attachment of relevant beliefs to nodes
and arcs. Recursive algorithms allow computation of belief transforms
for node adjustment, and propagation of updated means, variances and
covariances 
through the tree. 
Note that whilst propagating outwards through the tree, updating of
the different branches of the tree may proceed in parallel. \mupad\
provides a ``parallel for'' construct which allows simple exploitation
of this fact on appropriate hardware. 
Simple functions to allow computation of diagnostics
for particular nodes also exist. 


\section{Conclusions}
The algorithms described in this paper are very simple and easy to
implement, and very fast compared to many other algorithms for
updating in Bayesian belief networks. Further, by linking the theory
with the  machinery of the Bayes linear methodology, full \emph{a
  priori} and diagnostic analysis may also take place. \emph{A priori}
analysis is particularly important in large sparse networks, where it
is often not clear whether or not it is worth observing particular
nodes, which may be ``far'' from nodes of interest. Similarly,
diagnostic analysis is crucial, both for diagnosing misspecified
node and arc beliefs, and for diagnosing an incorrectly structured
model.

For those who already appreciate the benefits of working within
the Bayes linear paradigm, the methodology described in this paper
provides a mechanism for the tackling of much larger structured
problems than previously possible, using local computation of belief
transforms, adjustments and diagnostics.

\end{document}